\documentclass[runningheads]{llncs}
\usepackage[T1]{fontenc}
\usepackage{graphicx,verbatim}
\usepackage{amsmath} 
\usepackage{booktabs}
\usepackage{tabularx}
\usepackage{multirow}
\usepackage{amssymb}

\usepackage{subfigure}
\begin{document}
\title{Prediction of Frozen Region Growth in Kidney Cryoablation Interventoin Using a 3D Flow-Matching Model}
%
\begin{comment}  %% Removed for anonymized MICCAI 2025 submission
\author{First Author\inst{1}\orcidID{0000-1111-2222-3333} \and
Second Author\inst{2,3}\orcidID{1111-2222-3333-4444} \and
Third Author\inst{3}\orcidID{2222--3333-4444-5555}}
%
\authorrunning{F. Author et al.}
% First names are abbreviated in the running head.
% If there are more than two authors, 'et al.' is used.
%
\institute{Princeton University, Princeton NJ 08544, USA \and
Springer Heidelberg, Tiergartenstr. 17, 69121 Heidelberg, Germany
\email{lncs@springer.com}\\
\url{http://www.springer.com/gp/computer-science/lncs} \and
ABC Institute, Rupert-Karls-University Heidelberg, Heidelberg, Germany\\
\email{\{abc,lncs\}@uni-heidelberg.de}}

\end{comment}

\author{Siyeop Yoon\inst{1}
\and Yujin Oh \inst{1}
\and Matthew Tivnan\inst{1}
\and Sifan Song \inst{1}
\and Pengfei Jin\inst{1}
\and Sekeun Kim\inst{1}
\and Hyun Jin Cho\inst{2}
\and Dufan Wu \inst{1}
\and Raul Uppot \inst{1}
\and Quanzheng Li \inst{1\dag}}
\authorrunning{S. Yoon et al.}
% First names are abbreviated in the running head.
% If there are more than two authors, 'et al.' is used.
%

\institute{Massachusetts General Hospital and Harvard Medical School, Boston, MA, USA \and
Chungnam National University College of Medicine, South Korea\\
\email{li.quanzheng@mgh.harvard.edu}\\
}

\maketitle              % typeset the header of the contribution
\begin{abstract}
This study presents a 3D flow-matching model designed to predict the progression of the frozen region (iceball) during kidney cryoablation. Precise intraoperative guidance is critical in cryoablation to ensure complete tumor eradication while preserving adjacent healthy tissue. However, conventional methods, typically based on physics-driven or diffusion-based simulations, are computationally demanding and often struggle to accurately represent complex anatomical structures.

To address these limitations, our approach leverages intraoperative CT imaging to inform the model. The proposed 3D flow-matching model is trained to learn a continuous deformation field that maps early-stage CT scans to future predictions. This transformation not only estimates the volumetric expansion of the iceball but also generates corresponding segmentation masks, effectively capturing spatial and morphological changes over time.
Quantitative analysis highlights the model’s robustness, demonstrating strong agreement between predictions and ground-truth segmentations. The model achieves an Intersection over Union (IoU) score of 0.61 ± 0.11 and a Dice coefficient of 0.75 ± 0.11. By integrating real-time CT imaging with advanced deep learning techniques, this approach has the potential to enhance intraoperative guidance in kidney cryoablation, improving procedural outcomes and advancing the field of minimally invasive surgery.
% Authors must provide keywords and are not allowed to remove this Keyword section.

\end{abstract}

\section{Introduction}

Kidney cryoablation is a minimally invasive treatment for small renal tumors that effectively destroys the tumor while preserving kidney function \cite{stacul2021cryoablation}. In this procedure, cryoprobes are inserted percutaneously under CT guidance to create an “iceball,” a clearly defined area of frozen tissue that marks the ablation zone and protects nearby critical structures \cite{knox2020intermediate,zhou2019radiofrequency}. On CT images, the iceball appears as a low-attenuation (hypodense) region, usually with a spherical or ellipsoidal shape. Its well-defined borders are produced by ice crystal formation, which decreases the tissue density. Often, the medial edge of the iceball is shorter because the warmer blood flow in the central kidney limits the spread of freezing (the “cold sink” effect). Accurate, real-time prediction of the iceball’s growth is critical to ensure that the entire tumor is covered with an adequate safety margin, thereby minimizing injury to healthy tissue \cite{littrup2009lethal}.

Conventional strategies for predicting iceball dynamics have spanned from simple geometric approximations to sophisticated physics-based simulations, notably those employing the Pennes bioheat equation \cite{kim2007finite,tanwar2023numerical}. Although these methods can yield high accuracy, their substantial computational overhead and dependence on simplifying assumptions restrict their applicability in real-time intraoperative settings. Recent data-driven approaches, particularly those utilizing deep learning, have demonstrated promising improvements in prediction accuracy within analogous ablation contexts \cite{moreira2023ai}. Moreover, advances in artificial intelligence have transformed medical image analysis; early generative frameworks, such as Variational Autoencoders and Generative Adversarial Networks, laid the foundation for realistic image synthesis \cite{kingma2013auto,goodfellow2020generative}, while modern diffusion-based models have significantly enhanced image fidelity and control over the generative process \cite{song2020score,ho2020denoising,EDM}. Even with these advances, significant challenges still exist—particularly the high computational costs and long processing times required for full 3D volumes. Techniques such as patch-wise processing \cite{wang2023patch,yoon2024high} and working in latent space \cite{rombach2022high} have been suggested to ease these issues, but keeping the anatomical accuracy of the predictions remains difficult. 

In our study, we utilize the principles of flow-matching models \cite{lipman2022flow,lipman2024flow} to develop a novel 3D predictive framework for the future configuration of the ablation zone. Unlike conventional diffusion models that rely on iterative noise removal, our method directly learns a deformation field that continuously transforms early CT images into their predicted future states. This direct mapping preserves the high image quality and precise control characteristic of diffusion-based techniques, while significantly reducing computational requirements. Furthermore, by incorporating patch-wise training strategies with the usage of a residual estimation approach, our model achieves an effective balance between computational efficiency and the retention of spatial and anatomical detail. Notably, as most anatomical structures remain unchanged during the procedure—with only the iceball exhibiting dynamic changes—our approach is specifically tailored to capture the evolving deformation of the ablation zone.

\section{Methods}
\subsection{Data Description and Study Population}
This study utilized a retrospective dataset of 24 patients who underwent CT-guided percutaneous renal cryoablation between September 2017 and May 2023. The study was approved by the Institutional Review Board (IRB), and informed consent was waived due to the retrospective nature of the study (Protocol: Anonymized). Each procedure involved 2--4 freeze--thaw cycles, with intraoperative CT images acquired at regular intervals (approximately every 2--3 minutes) during each freeze to monitor iceball formation. The complete dataset comprises 31 imaging studies, partitioned into training and testing sets on a per-patient basis to avoid data leakage. For testing, 8 cases were included, yielding 34 target scans acquired from the 3-minute scan. Sixteen scans corresponded to the first cryoablation cycle, while 18 scans were obtained during the second cycle, with target scans acquired 6--10 minutes after the initial 3-minute scan. All patients underwent CT-guided percutaneous renal cryoablation using devices manufactured by various vendors.
\begin{figure}[t]
    \centering
    \includegraphics[width=0.85\linewidth]{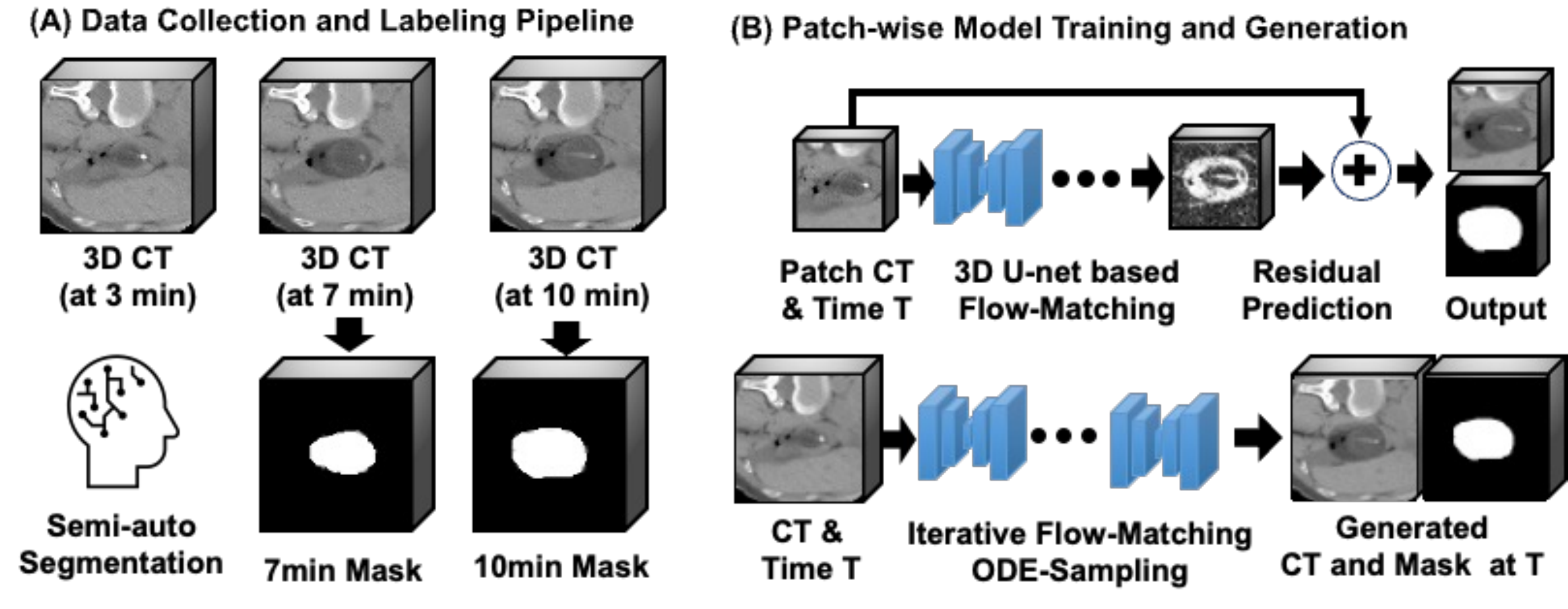}
    \caption{(A) Data collection and labeling pipeline, showing semi-automated segmentation of 3D CT scans acquired at 3-10 minutes to generate corresponding masks. (B) Patch-wise model training and generation framework, employing a 3D U-Net–based flow-matching approach with iterative ODE sampling to produce time-specific CT volumes and segmentation masks.}
    \label{fig:enter-label}
\end{figure}{}

\subsubsection{Model Inputs and Outputs}
The proposed model accepts a structured tuple; specifically, the inputs consist of the source volume \(X\), a stage indicator representing the current freeze cycle phase, and time change information (e.g., elapsed time or \(\Delta t\)). The trained model simultaneously predicts both the future CT image and the corresponding iceball segmentation, as shown in the Figure \ref{fig:enter-label}. The output is a tuple comprising the predicted target CT volume \(Y\) and a binary segmentation mask delineating the iceball. Joint optimization for image synthesis and segmentation ensures that the network effectively captures both the appearance and spatial extent of the evolving iceball.

\subsubsection{Flow-Matching for Residual Transformation}
To predict the future state of the frozen region, we employ a 3D flow-matching model that learns a continuous mapping from an early CT image, denoted by \(I_{\text{src}}\) (e.g., acquired at 3 minutes), to a predicted future CT image, denoted by \(I_{\text{tgt}}\) (e.g., acquired at 10 minutes), while simultaneously predicting the corresponding iceball segmentation at the future time point.

Let \(I_{\text{src}} \in \mathbb{R}^{H \times W \times D}\) denote the source CT volume and \(I_{\text{tgt}} \in \mathbb{R}^{H \times W \times D}\) denote the paired target volume. We define the residual volume as
\[
r = I_{\text{tgt}} - I_{\text{src}}.
\]
To model the transformation from \(I_{\text{src}}\) to \(I_{\text{tgt}}\), we introduce a normalized interpolation parameter \(\tau \in [0,1]\), such that the intermediate volume is given by
\[
I(\tau) = I_{\text{src}} + \tau\,r = I_{\text{src}} + \tau\,(I_{\text{tgt}} - I_{\text{src}}).
\]
Thus, \(I(0) = I_{\text{src}}\) and \(I(1) = I_{\text{tgt}}\). Because the interpolation is linear, the derivative with respect to \(\tau\) is constant, $ \frac{dI(\tau)}{d\tau} = r$

This implies that the true residual velocity field is $u(\tau) = r,\quad \forall\, \tau \in [0,1]$.
A neural network \(\mathcal{N}_\theta\) is employed to estimate the residual velocity field from the intermediate image \(I(\tau)\) and the interpolation parameter \(\tau\), yielding the prediction
\begin{equation}
\hat{u}(\tau) = \mathcal{N}_\theta\bigl(I(\tau), \tau\bigr).
\end{equation}
The training objective minimizes the mean squared error (MSE) between the predicted velocity field and the true residual:
\begin{equation}
    \mathcal{L}(\theta) = \mathbb{E}_{(I_{\text{src}},I_{\text{tgt}}),\, \tau \sim \mathcal{U}(0,1)} \left[ \left\| \mathcal{N}_\theta\Bigl(I_{\text{src}} + \tau\,(I_{\text{tgt}}-I_{\text{src}}), \tau\Bigr) - (I_{\text{tgt}}-I_{\text{src}}) \right\|^2 \right].
\end{equation}

To further increase data diversity and computational efficiency, the model is also trained in a patch-wise manner. Three-dimensional patches of size \(32 \times 32 \times 32\) are extracted from paired CT volumes and their corresponding iceball masks. Let \(P_{\text{src}}\) and \(P_{\text{tgt}}\) denote a pair of patches extracted from two temporally paired volumes. The residual patch is defined as
\begin{equation}
r_{\text{patch}} = P_{\text{tgt}} - P_{\text{src}} \text{, and }
P(\tau) = P_{\text{src}} + \tau\,r_{\text{patch}},\quad \tau \in [0,1],
\end{equation}
so that \(P(0) = P_{\text{src}}\) and \(P(1) = P_{\text{tgt}}\). Under this formulation, the derivative is constant:
\[
\frac{dP(\tau)}{d\tau} = r_{\text{patch}}.
\]
The patch-wise flow-matching loss is then given by
\begin{equation}
\label{patchloss}
\mathcal{L}_{\text{patch}}(\theta) = \frac{1}{N} \sum_{i=1}^{N} \left\| \mathcal{N}_\theta\bigl(P(\tau_i), \tau_i\bigr) - r_{\text{patch},i} \right\|^2,
\end{equation}
where \(N\) is the number of patch pairs in the batch and each \(\tau_i\) is sampled uniformly from \([0,1]\). The trained 3D U-Net is used to predict the future CT image by integrating the estimated velocity field over \(\tau\):
\begin{equation}
    I_{\text{tgt}} = I_{\text{src}} + \int_{0}^{1} \mathbf{v}(x,\tau) \, d\tau.
\end{equation}

This integration is performed using an ODE solver (e.g., the Heun 2nd order solver). In our implementation, the segmentation mask is provided as an additional input channel to the network. This enables simultaneous prediction of the future iceball segmentation; thus, the overall training objective is the patch-wise flow-matching loss calculated from 2-channel volume data. 

Finally, the 3D flow-matching model produces two outputs: a synthetic CT volume representing the kidney and the evolving iceball at a future time point, and a segmentation mask of the frozen region. We threshold segmentation channel with 0.95 to generate binary masks. Additionally, the framework allows estimation of the continuous growth of the iceball by controlling a stage indicator representing the current freeze cycle phase and time change information, which are provided as an additional input channel to the model.

\subsection{Data preprocessing and model implementation}
For the training, CT images were preprocessed to ensure consistent spatial representation and robust segmentation. First, all volumes were resampled to an isotropic resolution of 1 mm3. Next, within each freeze cycle, images were rigidly registered to compensate for respiratory motion using ITK\cite{yoo2002engineering}. The frozen region was segmented using ITK-SNAP \cite{yushkevich2016itk}: a semi-automatic approach yielded initial ground truth labels, which were further refined via manual segmentation to include the catheter. For optimal delineation, the CT visualization window was set to [-150, 350] HU. Finally, random 90,180,270 degrees rotations were applied during training to improve model generalization by considering various catheter insertion locations and directions during the procedure.

The training dataset was constructed by randomly pairing CT scans acquired at different time points within the same cryoablation cycle. For instance, scans acquired at 6 minutes were paired with those at 10 minutes (time difference = +4 minutes), and scans at 3 minutes (time difference = -3 minutes). This strategy augments the number of training pairs and introduces pairs with negative time differences, artificially reflecting a reverse cryoablation process.

The magnitude-preserving 3D U-Net was used as a residual velocity field estimator by minimizing patch-wise loss, Eq. \ref{patchloss}. We extended the convolutional layers of the EDM2 model to 3D and deactivated the preconditioning step used in the original implementation. The resulting model effectively preserves the magnitude of the feature representations throughout the network. 
Training was performed with the following hyperparameter settings (duration = 12k gradient update iterations, batch = 2048, channels = 64, learning rate = 0.0170, learning rate half-life = 35000 on 8 A100 GPUs over a 24-hour period). For comparative analysis, a diffusion model was also evaluated using the same settings with an EDM sampler employing 10-100 steps. The full model comprises approximately 80 million parameters and employs mixed-precision (FP16) training to accelerate computation. Under these conditions, both the predicted CT volume and the iceball mask are generated in approximately 15 seconds per inference on our hardware (single A100 GPU) with a cropped volume around the cryoablation catheter (size of $64^3$).

\subsection{Evaluation}
\begin{figure}[t]
    \centering
    \includegraphics[width=0.65\linewidth]{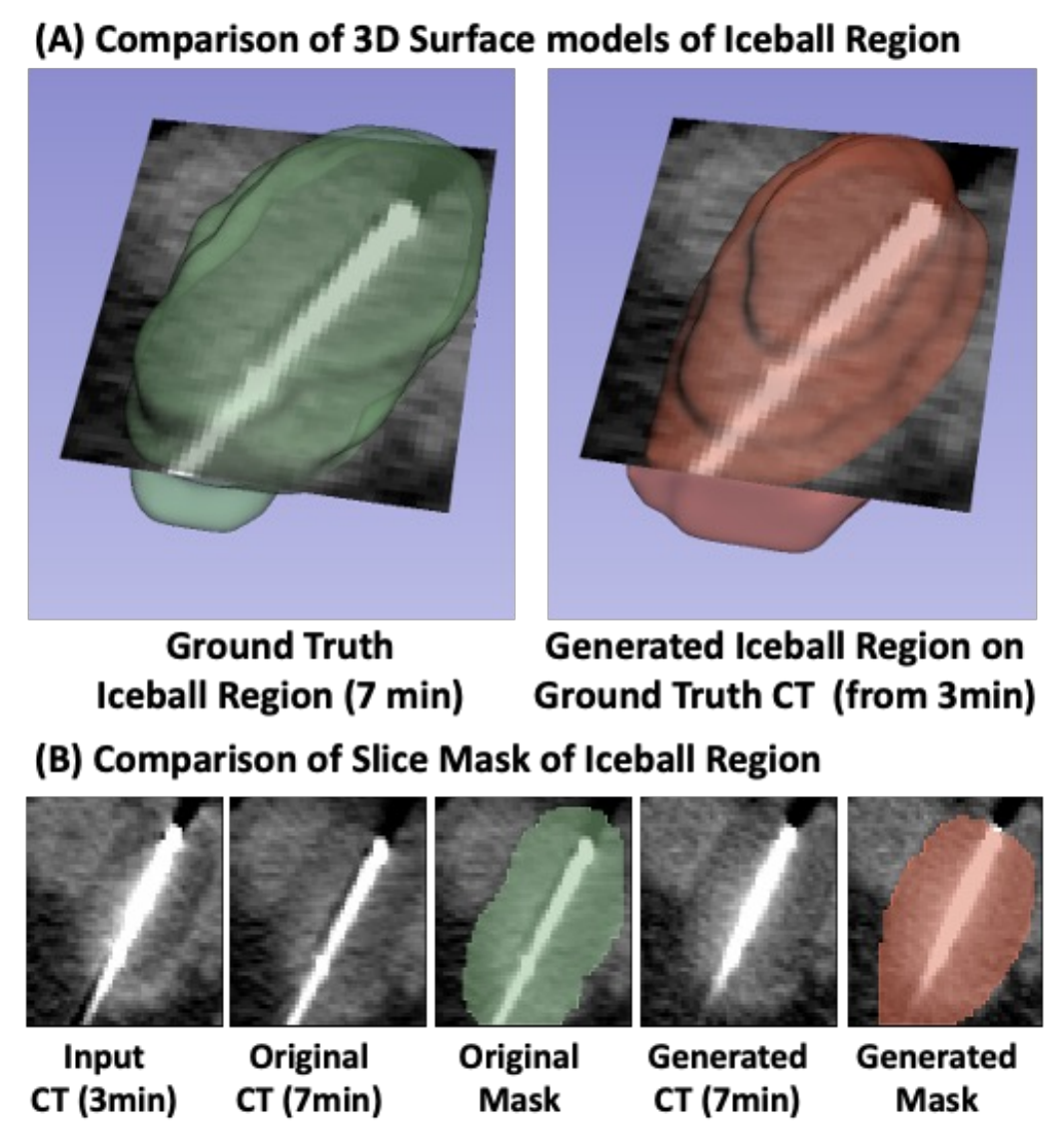}
    \caption{(A) Comparison of 3D surface models of the Iceball region, contrasting the ground truth (green) with the generated region (red) overlaid on axial CT slices. (B) Slice-by-slice comparison of the Iceball region masks, illustrating input and generated CTs alongside corresponding ground-truth and generated masks.}
    \label{fig:mainl}
\end{figure}{}
Our evaluation comprises both quantitative metrics and qualitative visualizations to assess the performance of the proposed flow-matching model for predicting iceball evolution. For quantitative evaluation, image quality was measured using standard metrics including normalized Mean Absolute Error (NMAE), Peak Signal-to-Noise Ratio (PSNR), and Structural Similarity Index (SSIM). These metrics were computed on CT images that were first normalized from the original range \([-150,350]\) Hounsfield Units (HU) to the range \([0,1]\). It should be noted that image-based evaluation metrics may not fully capture the performance of generative AI models, as such metrics can be sensitive to minor variations in texture and structural details that do not necessarily impact the clinical utility of the synthesized images. Therefore, quantitative metrics should be interpreted in conjunction with qualitative visual assessments. For segmentation performance, we calculated the Dice Similarity Coefficient (DSC), Intersection over Union (IoU), and the absolute volume difference between the predicted and ground truth masks.

\begin{table*}[t!]
\newcolumntype{C}{>{\centering\arraybackslash}X}
\centering
    \caption{Performance evaluation of Iceball Region and Segmentation Prediction.}
    \label{tab:results}
    \begin{tabularx}{\textwidth}{lCCCCCCC}\hline\hline
        \textbf{Method} &  NMAE (x10)   &  PSNR (dB) & SSIM  &IoU  & Dice\\\hline
        Diffusion (10) &  0.87 ± 0.2	&18.67 ± 2.19	&0.360 ± 0.14	&0.59 ± 0.11 &0.73 ± 0.09\\
        Diffusion (50) & 0.82 ± 0.2&	19.17 ± 2.40
	& 0.400 ± 0.15& 	0.56 ± 0.14	&	0.70 ± 0.13\\
        Diffusion (100) &0.83 ± 0.2&	19.13 ± 2.39&	0.403 ± 0.15&	0.55 ± 0.11&	0.71 ± 0.09  \\
        Flow  (10) &  0.82 ± 0.2	& 19.23 ± 2.35 &	0.403 ± 0.15	& 0.57 ± 0.13 &	0.72 ± 0.11\\
        Flow  (50) & 0.82 ± 0.2	&19.21 ± 2.48	&0.404 ± 0.15	&0.59 ± 0.11 &0.73 ± 0.10  \\
        Flow  (100) & 0.80 ± 0.2 & 19.32 ± 2.32 & 0.410 ± 0.15 &0.61 ± 0.11& 0.75 ± 0.11  \\
        
        \hline
    \end{tabularx} % Ensure column definitions and data match properly
\end{table*}

\section{Results}

The representative 3D surface comparison reveals that the generated Iceball region closely follows the contour and volume of the ground truth, showing minimal discrepancies in shape and boundary (Figure \ref{fig:mainl}). The slice-wise masks further confirm this accuracy, as the generated segmentation exhibits strong agreement with the original mask across multiple cross-sections. These findings indicate that the proposed method effectively captures both global and local features of the ablation zone, resulting in precise reconstructions of the Iceball region.

The quantitative evaluation of the Iceball region prediction and segmentation is summarized in Table\ref{tab:results}. Two methods—Diffusion and Flow models-were assessed across three sampling step settings (10, 50, and 100 iterations). For the Diffusion method, the performance at 10 iterations shows an NMAE of 0.87, a PSNR of 18.67dB, an SSIM of 0.360, an IoU of 0.59, and a Dice coefficient of 0.73. Increasing the iterations to 50 and 100 yields modest improvements in PSNR and SSIM, with values around 19.17dB/0.400 and 19.13dB/0.403 respectively. However, the IoU and Dice scores do not exhibit a consistent enhancement, indicating that higher iterations may refine certain image quality aspects without a proportional gain in segmentation accuracy.

In contrast, the Flow method demonstrates a clear trend of improvement with increased iterations. At 10 iterations, Flow achieves an NMAE of 0.82, a PSNR of 19.23dB, an SSIM of 0.403, an IoU of 0.57, and a Dice of 0.72. This performance further improves at 50 iterations and reaches its peak at 100 iterations—with an NMAE of 0.80, a PSNR of 19.32dB, an SSIM of 0.410, an IoU of 0.61, and a Dice coefficient of 0.75. These results indicate that the Flow-based approach is particularly effective in enhancing both region prediction accuracy and segmentation quality when provided with a higher number of iterations.

\begin{figure}[t]
    \centering
    \includegraphics[width=0.85\linewidth]{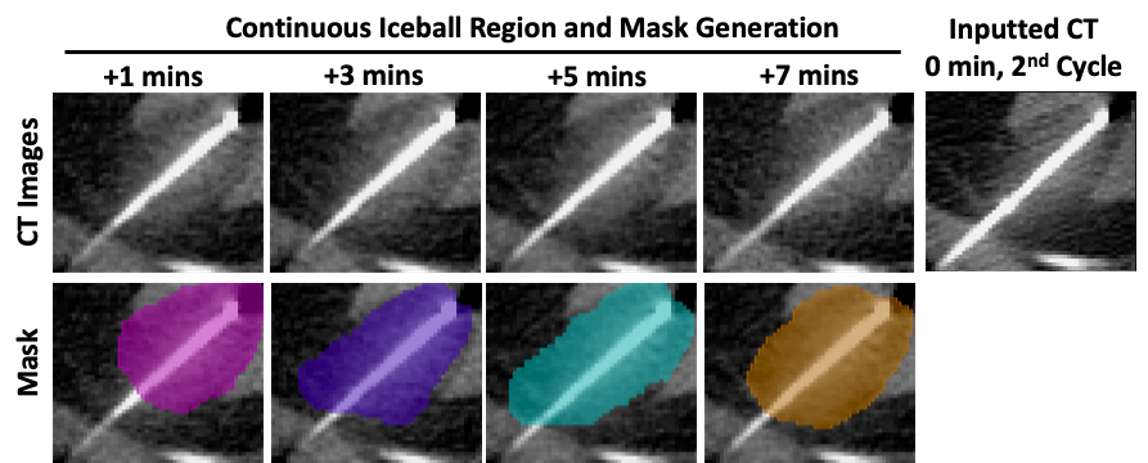}
    \caption{Continuous iceball region and mask generation for a second cryoablation cycle, starting from an input CT at 0 minutes. Each column depicts the predicted CT slice (top row) and its corresponding iceball mask (bottom row) at incremental time points of +1, +3, +5, and +7 minutes.}
    \label{fig:continuous_mapping}
\end{figure}
For qualitative evaluation, we synthesized CT images at two-minute intervals starting from the startign image at the second cryoablation cycle. These continuous mapping visualizations (see Figure~\ref{fig:continuous_mapping}) clearly illustrate the progressive evolution of the iceball. Each synthesized image shows incremental changes in the frozen region, capturing the subtle deformations and growth patterns over time. The smooth, gradual changes observed in the synthesized images and segmentation predictions highlights the model's capacity to effectively capture and predict the continuous evolution of the iceball.

\section{Discussion} 

This work introduces a 3D flow-matching model for predicting the progression of frozen regions in kidney cryoablation. By leveraging patch-wise training, the approach efficiently handles volumetric data while retaining spatial details. Comparative evaluations show that the flow-matching framework consistently outperforms a diffusion-based baseline in both image-quality metrics and segmentation accuracy, especially when employing higher iteration steps. Visual assessments confirm that the generated volumes and masks align closely with ground-truth . 

The proposed 3D flow-matching model offers a data-driven approach to predicting the spatial evolution of frozen regions in kidney cryoablation. By learning a continuous deformation field that connects early and later time points, the method directly models local changes in the ablation zone without requiring complex physics-based simulations or iterative noise-removal steps. This design choice leverages both patch-wise training and an ODE solver to maintain computational feasibility, making it adaptable to the time-sensitive demands of intraoperative settings.

Nevertheless, a few limitations should be noted. First, the retrospective dataset, although diverse in terms of imaging intervals and tumor sizes, is drawn from a single institution. As a result, generalization to other clinical contexts—such as different scanner settings or patient populations—remains to be fully explored.  Future research could extend this flow-matching framework by incorporating additional physiological variables (e.g., blood perfusion or tissue thermal properties) and exploring multi-task learning setups that predict not only the iceball boundary but also associated tissue viability or necrosis. Real-time or near-real-time inference can be further facilitated by optimizing network architectures and parallelizing the ODE-solver steps. Finally, prospective validation on larger, multicenter datasets is warranted to confirm the clinical utility and robustness of this approach. 
\bibliographystyle{splncs04}
\bibliography{Paper}

\begin{thebibliography}{10}
\providecommand{\url}[1]{\texttt{#1}}
\providecommand{\urlprefix}{URL }
\providecommand{\doi}[1]{https://doi.org/#1}

\bibitem{goodfellow2020generative}
Goodfellow, I., Pouget-Abadie, J., Mirza, M., Xu, B., Warde-Farley, D., Ozair, S., Courville, A., Bengio, Y.: Generative adversarial networks. Communications of the ACM  \textbf{63}(11),  139--144 (2020)

\bibitem{ho2020denoising}
Ho, J., Jain, A., Abbeel, P.: Denoising diffusion probabilistic models. Advances in neural information processing systems  \textbf{33},  6840--6851 (2020)

\bibitem{EDM}
Karras, T., Aittala, M., Aila, T., Laine, S.: Elucidating the design space of diffusion-based generative models. Advances in Neural Information Processing Systems  \textbf{35},  26565--26577 (2022)

\bibitem{kim2007finite}
Kim, C., O'Rourke, A.P., Mahvi, D.M., Webster, J.G.: Finite-element analysis of ex vivo and in vivo hepatic cryoablation. IEEE transactions on biomedical engineering  \textbf{54}(7),  1177--1185 (2007)

\bibitem{kingma2013auto}
Kingma, D.P., Welling, M., et~al.: Auto-encoding variational bayes (2013)

\bibitem{knox2020intermediate}
Knox, J., Kohlbrenner, R., Kolli, K., Fidelman, N., Kohi, M.P., Lehrman, E., Lokken, R.P., Zagoria, R., Kerlan, R.K., LaBerge, J., et~al.: Intermediate to long-term clinical outcomes of percutaneous cryoablation for renal masses. Journal of Vascular and Interventional Radiology  \textbf{31}(8),  1242--1248 (2020)

\bibitem{lipman2022flow}
Lipman, Y., Chen, R.T., Ben-Hamu, H., Nickel, M., Le, M.: Flow matching for generative modeling. arXiv preprint arXiv:2210.02747  (2022)

\bibitem{lipman2024flow}
Lipman, Y., Havasi, M., Holderrieth, P., Shaul, N., Le, M., Karrer, B., Chen, R.T., Lopez-Paz, D., Ben-Hamu, H., Gat, I.: Flow matching guide and code. arXiv preprint arXiv:2412.06264  (2024)

\bibitem{littrup2009lethal}
Littrup, P.J., Jallad, B., Vorugu, V., Littrup, G., Currier, B., George, M., Herring, D.: Lethal isotherms of cryoablation in a phantom study: effects of heat load, probe size, and number. Journal of Vascular and Interventional Radiology  \textbf{20}(10),  1343--1351 (2009)

\bibitem{moreira2023ai}
Moreira, P., Tuncali, K., Tempany, C., Tokuda, J.: Ai-based isotherm prediction for focal cryoablation of prostate cancer. Academic radiology  \textbf{30},  S14--S20 (2023)

\bibitem{rombach2022high}
Rombach, R., Blattmann, A., Lorenz, D., Esser, P., Ommer, B.: High-resolution image synthesis with latent diffusion models. In: Proceedings of the IEEE/CVF conference on computer vision and pattern recognition. pp. 10684--10695 (2022)

\bibitem{song2020score}
Song, Y., Sohl-Dickstein, J., Kingma, D.P., Kumar, A., Ermon, S., Poole, B.: Score-based generative modeling through stochastic differential equations. arXiv preprint arXiv:2011.13456  (2020)

\bibitem{stacul2021cryoablation}
Stacul, F., Sachs, C., Giudici, F., Bertolotto, M., Rizzo, M., Pavan, N., Balestreri, L., Lenardon, O., Pinzani, A., Pola, L., et~al.: Cryoablation of renal tumors: long-term follow-up from a multicenter experience. Abdominal Radiology  \textbf{46},  4476--4488 (2021)

\bibitem{tanwar2023numerical}
Tanwar, S., Famhawite, L., Verma, P.R.: Numerical simulation of bio-heat transfer for cryoablation of regularly shaped tumours in liver tissue using multiprobes. Journal of Thermal Biology  \textbf{113},  103531 (2023)

\bibitem{wang2023patch}
Wang, Z., Jiang, Y., Zheng, H., Wang, P., He, P., Wang, Z., Chen, W., Zhou, M.: Patch diffusion: Faster and more data-efficient training of diffusion models. arXiv preprint arXiv:2304.12526  (2023)

\bibitem{yoo2002engineering}
Yoo, T.S., Ackerman, M.J., Lorensen, W.E., Schroeder, W., Chalana, V., Aylward, S., Metaxas, D., Whitaker, R.: Engineering and algorithm design for an image processing api: a technical report on itk-the insight toolkit. In: Medicine Meets Virtual Reality 02/10, pp. 586--592. IOS press (2002)

\bibitem{yoon2024high}
Yoon, S., Pratap, J.S., Liu, W.C., Tivnan, M., Ren, H., Bhashyam, A., Li, Q., Chen, N., Li, X.: High-resolution 3d ct synthesis from bidirectional x-ray images using 3d diffusion model. In: 2024 IEEE International Symposium on Biomedical Imaging (ISBI). pp.~1--4. IEEE (2024)

\bibitem{yushkevich2016itk}
Yushkevich, P.A., Gao, Y., Gerig, G.: Itk-snap: An interactive tool for semi-automatic segmentation of multi-modality biomedical images. In: 2016 38th annual international conference of the IEEE engineering in medicine and biology society (EMBC). pp. 3342--3345. IEEE (2016)

\bibitem{zhou2019radiofrequency}
Zhou, W., Herwald, S.E., McCarthy, C., Uppot, R.N., Arellano, R.S.: Radiofrequency ablation, cryoablation, and microwave ablation for t1a renal cell carcinoma: a comparative evaluation of therapeutic and renal function outcomes. Journal of Vascular and Interventional Radiology  \textbf{30}(7),  1035--1042 (2019)

\end{thebibliography}

\end{document}